\documentclass[12pt, bezier]{article}

\def\beq{\begin{equation}}
\def\eeq{\end{equation}}
\def\beqa{\begin{eqnarray}}
\def\eeqa{\end{eqnarray}}
\newtheorem{Lem}{Lemma}
\newtheorem{Th}{Theorem}

\newtheorem{Cor}{Corollary}

\begin{document}

\def\supp{\mathop{\rm supp}}
\def\diam{\mathop{\rm diam}}
\def\Exp{\mathop{\rm Exp}}
\def\Log{\mathop{\rm Log}}

\begin{titlepage}


\begingroup
\def\thirteen{\large\bf}
\def\ten{\bf}
\def\eight{\footnotesize}
\baselineskip 15pt
\thispagestyle{empty}

\centerline{\ten Centre de Physique
Th\'eorique\footnote
{\eight Unit\'e Propre de Recherche
7061}, CNRS Luminy, Case 907}

\centerline{\ten F-13288 Marseille -- Cedex 9}

\vskip 2truecm

\centerline{\thirteen A LECTURE ON CLUSTER EXPANSIONS}

\bigskip

\centerline{{\bf 
Salvador MIRACLE-SOL\'E\footnote
{\eight 
Centre de Physique
Th\'eorique, CNRS, Marseille}
}}

\vskip 2truecm


\begin{abstract}
A short exposition with complete proofs of the 
theory of cluster expansions for an abstract polymer system
is presented.

\end{abstract}

\bigskip

Cours au DEA de Physique Th\'eorique de Marseille. 
Contribution to the conference
{\it Phasen\"uberg\"ange}, 
Mathematisches Forschungsinstitut Oberwolfach, 
April 29th -- May 5th, 2001. 
This version appeared in 
{\it Markov Processes Relat.\  Fields} {\bf 16}, 287--294 (2010)


\bigskip 



\bigskip
\bigskip

\noindent Avril 2000

\noindent CPT-2000/P. 4115

\bigskip

\noindent anonymous ftp : ftp.cpt.univ-mrs.fr

\noindent web : www.cpt.univ-mrs.fr

\endgroup

\thispagestyle{empty}
\end{titlepage}



Consider a countable set ${\cal P}$, 
whose elements will be called {polymers}.
Let ${\cal I}$ be a subset of 
${\rm P}_2({\cal P})$, the set of all subsets of ${\cal P}$
with two elements.
We say that two polymers $\gamma$ and $\gamma'$ are
{incompatible} if $\{\gamma,\gamma'\}\in{\cal I}$ or 
if $\gamma=\gamma'$,
and we will also write $\gamma\not\sim\gamma'$.
If $\{\gamma,\gamma'\}\not\in{\cal I}$ we say that the two
polymers are {compatible} and we write $\gamma\sim\gamma'$. 

Assume that a complex valued function $\phi(\gamma)$,
$\gamma\in{\cal P}$, is given.
We call $\phi(\gamma)$ the {weight}, or the {activity},
of the polymer $\gamma$.
For any finite subset $\Lambda\subset{\cal P}$,
the {partition function} $Z(\Lambda)$ of the
polymer system is defined by 
\beq
Z(\Lambda)=
\sum_{{\scriptstyle X\subset\Lambda}
\atop{\scriptstyle {\rm compatible}}}
\prod_{\gamma\in X}\phi(\gamma)
\label{1}\eeq
The sum runs over all subsets $X$ of ${\Lambda}$ such that
$\gamma\sim\gamma'$ for any two distinct elements of $X$. 
If $X$ contains only one element, $X$ is considered a
compatible subset, and if $X=\emptyset$,
the product is interpreted as the number $1$.

We introduce the following
function on ${\cal P}\times{\cal P}$
\beq
f(\gamma,\gamma')=\cases{-1 &if $\gamma\not\sim\gamma'$ 
or $\gamma=\gamma'$\cr
0 &otherwise \cr}
\label{2}\eeq

Let ${\cal G}_n$, $n\ge2$ be the set of connected  
graphs with $n$ vertices, $1,\dots,n$.
We consider undirected graphs without multiple edges,  
equivalently defined by a subset of ${\rm P}_2(\{1,\dots,n\})$
which determines the edges.
Given $g\in{\cal G}_n$ we define the value of $g$ on a
sequence $(\gamma_1,\dots,\gamma_n)\in{\cal P}^n$ as
\beq
g(\gamma_1,\dots,\gamma_n)=\prod_{(i,j)\in g}f(\gamma_i,\gamma_j)
\label{3}\eeq
where $(i,j)\in g$ means that the graph $g$ has
an edge connecting $i$ with $j$.
We also define ${\cal G}_1$
as the set containing only one graph $g$ having
only one vertex (and no edges) and write
\beq
g(\gamma)=1,\quad \gamma\in{\cal P}
\label{4}\eeq

\begin{Th}[Expansion]
\label{T1}
Define
\beq
a^{\rm T}(\gamma_1,\dots,\gamma_{n})=
\sum_{g\in {\cal G}_{n}}g(\gamma_1,\dots,\gamma_{n})
\label{5}\eeq
Then, we have
\beq
\ln Z(\Lambda)=\sum_{n=1}^\infty {1\over{n !}}
\sum_{(\gamma_1,\dots,\gamma_{n})\in\Lambda^n}
a^{\rm T}(\gamma_1,\dots,\gamma_{n})
\prod_{i=1}^n\phi(\gamma_i)
\label{6}\eeq
\end{Th}
{\it Proof.} 
The partition function can be written as
\beq
Z(\Lambda)=1+\sum_{\gamma\in{\Lambda}}\phi(\gamma)+
\sum_{n=2}^\infty{1\over{n!}}
\sum_{(\gamma_1,\dots,\gamma_n)\in{\Lambda}^n}
\prod_{i=1}^n\phi(\gamma_i)
\prod_{1\le i<j\le n}\big(f(\gamma_i,\gamma_j)+1\big)
\label{7}\eeq
and developing the second product
\beqa
Z(\Lambda) &=&
1+\sum_{n=1}^\infty{1\over{n!}}\sum_{m=1}^\infty{1\over{m!}}
\sum_{{\scriptstyle k_1,\dots,k_m}\atop{\scriptstyle\sum_ik_i=n}}
\sum_{g_1\in {\cal G}_{k_1}}\sum_{g_2\in {\cal G}_{k_2}}\dots
\sum_{g_m\in {\cal G}_{k_m}}{{n!}\over{k_1!k_2!\dots k_m!}} \cr
&&\sum_{(\gamma_1,\dots,\gamma_n)\in\Lambda^n}  
g_1(\gamma_1,\dots,\gamma_{k_1}) 
g_2(\gamma_{k_1+1},\dots,\gamma_{k_1+k_2})\dots \cr
&&\kern2cm\dots g_m(\gamma_{k_1+\dots+k_{m-1}+1},\dots,\gamma_n)
\prod_{i=1}^n \phi(\gamma_i) \cr
&=& 1+\sum_{m=1}^\infty{1\over{m!}}\Bigg(
\sum_{k=1}^\infty{1\over{k !}}\sum_{g\in {\cal G}_{k}}
\sum_{(\gamma_1,\dots,\gamma_k)\in\Lambda^k}
g(\gamma_1,\dots,\gamma_{k})
\prod_{i=1}^k \phi(\gamma_i)\Bigg)^m \cr
&=& \exp \Bigg(
\sum_{k=1}^\infty{1\over{k !}}\sum_{g\in {\cal G}_{k}}
\sum_{(\gamma_1,\dots,\gamma_k)\in\Lambda^k}
g(\gamma_1,\dots,\gamma_{k})
\prod_{i=1}^k \phi(\gamma_i)\Bigg)
\label{8}\eeqa
The theorem is proved.

\bigskip

With any finite sequence $\Gamma=(\gamma_1,\dots,\gamma_{n})
\in{\cal P}^n$ we associate the graph $\theta(\Gamma)$
with vertices $\{1,\dots,n\}$ obtained by drawing an edge between
the vertices $i$ and $j$ if $\gamma_{i}\not\sim\gamma_{j}$ 
and also if $\gamma_{i}=\gamma_{j}$.
We observe that $g(\Gamma)\not=0$
only if $\theta(\Gamma)$ is a connected graph 
(i.e., if $\theta(\Gamma)\in{\cal G}_n$) and 
$g\in{\cal G}_n$ is a subgraph of $\theta(\Gamma)$.
Then
\beq
a^{\rm T}(\Gamma)
=\sum_{g\in{\cal G}_{n}}g(\Gamma)
=\sum_{{\scriptstyle g\subset\theta(\Gamma)}\atop
{\scriptstyle g\in{\cal G}_n}}(-1)^{\vert g\vert}
\label{9}\eeq
where $\vert g\vert$ is the number of edges of the 
graph $g$.
In other words, $a^{\rm T}(\Gamma)$ is equal to 
the number of connected subgraphs of $\theta(\Gamma)$ 
with an even number of edges minus the number of 
connected subgraphs with an odd number of edges.
If the graph $\theta(\Gamma)$ is not connected, 
$a^{\rm T}(\Gamma)=0$.

We observe that the $a^{\rm T}(\gamma_1,\dots,\gamma_n)$, 
$n=1,2,\dots$, are symmetric functions. 
Thus, we can write the considered expansions also 
in terms of multi-indices instead of finite sequences.
A multi-index $X$ on a set ${\cal P}$ is a function 
$X(\gamma)$, $\gamma\in{\cal P}$, taking non-negative 
integer values and such that 
$\supp X=\{\gamma\in{\cal P} : X(\gamma)\ge1\}$ 
is a finite set or, equivalently, such that 
$\vert X\vert=\sum_{\gamma\in{\cal P}}X(\gamma)$ is a
finite number. 
We denote by ${\rm M}(\Lambda)$ the set of 
multi-indices defined on $\Lambda$.

If $X$ is a given multi-index and 
$\Gamma=(\gamma_1,\dots,\gamma_{n})$, $n=\vert X\vert$, 
one of the sequences corresponding to $X$, we define
$a^{\rm T}(X)=a^{\rm T}(\Gamma)$, and a new function
\beq
\phi^{\rm T}(X)=
\Big(\prod_{\gamma\in{\cal P}}X(\gamma)!\Big)^{-1}
a^{\rm T}(X)\prod_{\gamma\in{\cal P}}\phi(\gamma)^{X(\gamma)}
\label{10}\eeq
Taking into account that
the number of different sequences $\Gamma$ associated
to $X$ is $n!/\prod_{\gamma\in{\cal P}}X(\gamma)!$,   
the statement in theorem \ref{T1} can be written as
\beq
\ln Z(\Lambda)=\sum_{X\in{\rm M}(\Lambda)}\phi^{\rm T}(X)
\label{11}\eeq
The multi-indices such that $a^{\rm T}(X)\ne0$ (i.e.,  
whose associated graph $\theta(\Gamma)$ is connected)
will be called {clusters.}

\begin{Th}[Convergence]
\label{T2}
Assume that there is a positive function $\mu(\gamma)$, 
$\gamma\in{\cal P}$, such that, for all $\gamma_0\in{\cal P}$, 
\beq
\vert\phi(\gamma_0)\vert\le\big(e^{\mu(\gamma_0)}-1\big)
\exp \bigg(-\sum_{\gamma\not\sim\gamma_0}\mu(\gamma)\bigg)
\label{i1}\eeq
Then, for all $\gamma_1\in{\cal P}$, we have
\beqa
\sum_{X\in {\rm M}({\cal P}),\,X(\gamma_1)\ge1}
\vert \phi^{\rm T}(X)\vert
&\le& \mu(\gamma_1)
\label{i2} \\ 
\sum_{X\in {\rm M}({\cal P})}
X(\gamma_1)\vert\phi^{\rm T}(X)\vert
&\le& e^{\mu(\gamma_1)}-1
\label{i3}
\eeqa
\end{Th}
We first prove the following lemma. 

\begin{Lem}
For any $X\in{\rm M}({\cal P})$, we have
\label{L1}
\beq
(-1)^{\vert X\vert+1}a^{\rm T}(X)\ge0 
\label{i4}\eeq
\end{Lem}

\noindent
{\it Proof of lemma \ref{L1}.}
We introduce the partition function
\beq
Z^*(\Lambda)= \sum_{{\scriptstyle X\subset\Lambda}
\atop{\scriptstyle {\rm compatible}}}
\prod_{\gamma\in X}\Big(-\vert\phi(\gamma)\vert\Big)
\label{15}\eeq
for which
\beq
-\ln Z^*(\Lambda) = \sum_{X\in{\rm M}(\Lambda)}
-\Big(\prod_{\gamma\in{\cal P}}X(\gamma)!\Big)^{-1}
a^{\rm T}(X)\prod_{\gamma}
\Big(-\vert\phi(\gamma)\vert\Big)^{X(\gamma)} 
\eeq
If inequality (\ref{i4}) is satisfied, we have
\beq
-\ln Z^*(\Lambda) =  
\sum_{X\in{\rm M}(\Lambda)}\vert\phi^{\rm T}(X)\vert
\label{16}\eeq
showing that lemma \ref{L1} is equivalent to the fact
that all terms in the expansion of $-\ln Z^*(\Lambda)$
are positive.
This fact will be proved by an induction argument 
on the subsets $\Lambda$. 
It certainly holds when $\Lambda$ contains only one polymer.
Assume that it holds for a given $\Lambda$ and
let $\gamma_0\in{\cal P}\backslash\Lambda$.
From the definition of $Z^*$ we see that
\beq
Z^*(\Lambda\cup\{\gamma_0\}) = Z^*(\Lambda)  
-\vert\phi(\gamma_0)\vert Z^*(\Lambda_0) 
\label{17}\eeq
with
$\Lambda_0=\{\gamma\in\Lambda : \gamma\sim\gamma_0\}$,
and  
\beq
-\ln\,Z^*(\Lambda\cup\{\gamma_0\}) = -\ln\,Z^*(\Lambda)  
-\ln \Bigg(1-{{\vert\phi(\gamma_0)\vert Z^*(\Lambda_0)}
\over{Z^*(\Lambda)}}\Bigg) 
\label{18}\eeq
On the other hand, we have
\beq
Z^*(\Lambda_0)/Z^*(\Lambda)=
\exp\sum_{X\in{\rm M}(\Lambda)\backslash{\rm M}(\Lambda_0)}
\vert\phi^{\rm T}(X)\vert
\label{19}\eeq
This shows the positivity of all the terms in the expansion of
the second term in the right hand side of equation (\ref{18}) 
(remark that the series expansions of the functions $\exp x$
and $-\ln(1-x)$ have only positive terms for $x\ge0$). 
Since, by assumption, this is also 
the case for the first term, it follows that also 
$-\ln\,Z^*(\Lambda\cup\{\gamma_0\})$ satisfies
the induction hypothesis. 
The lemma is proved.

\bigskip

\noindent{\it Proof of theorem \ref{T2}.}
We use again an induction argument on the subsets $\Lambda$.
Assume that, for a given $\Lambda$ and any $\gamma\in\Lambda$,
the following estimate holds 
\beq
\sum_{X\in{\rm M}(\Lambda),\,X(\gamma)\ge1}
\vert\phi^{\rm T}(X)\vert\le\mu(\gamma)
\label{20}\eeq
This inequality can also be written as 
\beq
-\ln Z^*(\Lambda)+\ln Z^*(\Lambda\backslash\{\gamma\})
\le\mu(\gamma)
\label{21}\eeq
and, for all $\Lambda'\subset\Lambda$, it implies 
\beq
-\ln Z^*(\Lambda)+\ln Z^*(\Lambda')
\le\sum_{\gamma\in\Lambda\backslash\Lambda'}\mu(\gamma)
\label{22}\eeq
and, in particular, 
\beq
{Z^*(\Lambda_0)}/{Z^*(\Lambda)}
\le \exp\bigg(
\sum_{\gamma:\gamma\in\Lambda, \gamma\not\sim\gamma_0} 
\mu(\gamma)\bigg)
\label{23}\eeq
because
$\Lambda\backslash\Lambda_0
=\{\gamma\in\Lambda : \gamma\not\sim\gamma_0\}$. 
Since $\Lambda$ does not contain $\gamma_0$, we get
\beq
\vert\phi(\gamma_0)\vert
{{Z^*(\Lambda_0)}\over{Z^*(\Lambda)}}
\le \vert\phi(\gamma_0)\vert \exp\bigg(-\mu(\gamma_0)
+\sum_{\gamma:\gamma\in{\cal P}, \gamma\not\sim\gamma_0} 
\mu(\gamma)\bigg)
\label{24}\eeq
and, taking the assumption (\ref{i1}) 
of the theorem into account, 
\beq
\bigg\vert{{\phi(\gamma_0)Z^*(\Lambda_0)}\over{Z^*(\Lambda)}}\bigg\vert
\le e^{-\mu(\gamma_0)}(e^{\mu(\gamma_0)}-1) 
= 1-e^{-\mu(\gamma_0)}
\label{25}\eeq
Then, using (\ref{18}) and the fact that  
$-\ln(1-x)$ is an increasing function of $x$, 
for any real $x$ in the interval $-1<x<1$, 
we obtain
\beq
-\ln{{Z^*(\Lambda\cup\{\gamma_0\})}\over{Z^*(\Lambda)}} 
\le -\ln \Big(1-\Big(1-e^{-\mu(\gamma_0)}\Big)\Big)
=\mu(\gamma_0)
\label{26}\eeq
This proves the induction hypothesis (\ref{20}) 
for $\Lambda\cup\{\gamma_0\}$, and therefore for all $\Lambda$  
(being valid when $\Lambda$ contains only one element). 
Statement (\ref{i2}) of the theorem is proved.

Finally, if $\gamma_0\in\Lambda$ and  
$\Lambda_0$ is defined as above, we have
\beq
\sum_{X\in{\rm M}(\Lambda)}
X(\gamma_0)\vert\phi^{\rm T}(X)\vert
= -\vert\phi(\gamma_0)\vert{\partial\over{\partial
\vert\phi(\gamma_0)\vert}}\ln Z^*(\Lambda)
={{\vert\phi(\gamma_0)\vert Z^*(\Lambda_0)}\over{Z^*(\Lambda)}}
\eeq
Then, the statement (\ref{i3}) follows from hypothesis 
(\ref{i1}), taking into account inequality (\ref{22}) 
and that now $\Lambda$ contains $\gamma_0$.
This ends the proof of the theorem.

\medskip

The following consequences of the theorem concern
the correlation functions and the truncated correlation 
functions.  
Notice that, for $\Lambda$ finite and
$\{\gamma_1,\dots,\gamma_n\}\subset\Lambda$, 
these functions can be written as
\beqa
\rho_\Lambda(\gamma_1,\dots,\gamma_n)
&=& 
Z(\Lambda)^{-1}\Big(\prod_{i=1}^n\phi(\gamma_i)\Big) 
Z(\cap_{i=1}^n\Lambda_i) 
\label{30} \\
&=& Z(\Lambda)^{-1}
\Big(\prod_{i=1}^n\big(\phi(\gamma_i)\,{\partial}/
{\partial\phi(\gamma_i)}\big)\Big) Z(\Lambda) \label{31}\\
\rho^{\rm T}_\Lambda(\gamma_1,\dots,\gamma_n) 
&=& \Big(\prod_{i=1}^n\big(\phi(\gamma_i)\,{\partial}/
{\partial\phi(\gamma_i)}\big)\Big)\ln Z(\Lambda)
\label{32} 
\eeqa
with 
$\Lambda_i=\{\gamma\in\Lambda : \gamma\sim\gamma_i\}$. 

\begin{Cor}
Under the hypothesis of theorem \ref{T2},
the thermodynamic limits ($\Lambda\!\uparrow\!{\cal P}$) of 
the correlation functions and the truncated correlation 
functions, $\rho$ and $\rho^{\rm T}$, exist.
Moreover
\beq
\vert\rho(\gamma_1,\dots,\gamma_n)\vert
\le\prod_{i=1}^n \big(e^{\mu(\gamma_i)}-1\big)
\eeq
\end{Cor}
{\it Proof.}
Expressions (\ref{30}), (\ref{31}) and (\ref{32}) show
that the corresponding expansions in terms of the 
$\phi^{\rm T}(X)$ are
\beqa
\rho_\Lambda(\gamma_1,\dots,\gamma_n)
&=& \Big(\prod_{i=1}^n\phi(\gamma_i)\Big) 
\exp\bigg(\sum_{X\in}\phi^{\rm T}(X)\bigg) \\
\rho^{\rm T}(\gamma_1,\dots,\gamma_n) 
&=& \sum_{X\in{\rm M}({\cal P})}
X(\gamma_1)\dots X(\gamma_n)\phi^{\rm T}(X)
\eeqa
where $Q(\gamma_1,\dots,\gamma_n)={\rm M}({\cal P})\backslash
{\rm M}(\cap_{i=1}^n{\cal P}_i)$ is the set of $X$
whose support contains a $\gamma$ incompatible with
one of the $\gamma_i$, $i=1,\dots,n$.
The corollary follows from the convergence of these
expansions stated in theorem \ref{T2}.
A simple extension of the argument leading  to 
inequality (\ref{i4}) proves the inequality stated in the 
corollary.

\begin{Cor}
Assume that, for all $\gamma_0\in{\cal P}$, 
\beq
\vert\phi(\gamma_0)\vert\le e^{-t}\big(e^{\mu(\gamma_0)}-1\big)
\exp \bigg(-\sum_{\gamma\not\sim\gamma_0}\mu(\gamma)\bigg)
\label{36}\eeq
where $e^{-t}<1$ is a uniform factor.
Then the following cluster property is satisfied
(for all $\gamma_1\in{\cal P}$)
\beq
\sum_{(\gamma_2,\dots,\gamma_n)\in{\cal P}^{n-1}}
\vert\rho^{\rm T}_\Lambda(\gamma_1,\dots,\gamma_n)\vert
\le (n-1)!\,t^{-n+1}\big(e^{\mu(\gamma_1)}-1\big) 
\label{37}
\eeq
\end{Cor}
{\it Proof.}
Theorem \ref{T2} and assumption (\ref{36}) imply the bound
\beq
\sum_{X\in{\rm M}({\cal P})}
X(\gamma_1)e^{t\,\vert X\vert}\vert\phi^{\rm T}(X)\vert
\le \big(e^{\mu(\gamma_1)}-1\big)
\eeq
Taking into account equation (\ref{32}) and that
\beq
e^{t\,\vert X\vert}=
\sum_{n=0}^\infty {{t^n}\over{n!}} 
\bigg(\sum_{\gamma\in{\cal P}}X(\gamma)\bigg)^n
\eeq
we obtain the stated estimate (\ref{37}) from this bound.

\bigskip
\bigskip

\noindent{\Large\bf Bibliographical note}

\bigskip
\bigskip

The rigorous study of perturbation series in 
statistical mechanics has quite a long history
since the times when Lebowitz and Penrose
proved the convergence of the virial expansion \cite{1}, 
or the earlier result by Groeneveld \cite{0} in the case of  
repulsive potentials (using an alternating sign property 
similar to that of Lemma 1).
Other related early works are described in \cite{2},\cite{3}. 
In this context, polymer systems and the corresponding
expansions, also called cluster expansions,
have a particular interest
as they permitted to analyze, among other questions,
high and low temperature properties in
the case of lattice systems (see, for instance,
\cite{4} chapter V). 
There are several approaches to the proof of 
the convergence of cluster expansions:  
one is based on the use of Kirkwood-Salsburg
type of equations \cite{5},\cite{6}, 
others in bounding each term of the series by 
some combinatorics of trees on a graph \cite{7},\cite{8},  
\cite{4} (chapter V). 
In \cite{9} such bounds were obtained from a 
recurrence relation similar to the one 
used here in the first part of the proof, 
and in \cite{10} the M\"obius inversion formula
was used. 
In a recent work \cite{11}, Dobrushin used an induction
argument to obtain uniform bounds on the partition
functions and, as a consequence, analyticity 
properties of the system.

The proof of the convergence of cluster expansions 
presented in this note, and in ref.\ \cite{12}, is
based in part on the last mentioned work.  
This proof is in our view rather simple and direct. 
The result, stated in Theorem 2,   
is expressed in the ``classical'' form, 
useful in the applications, that is 
as a bound on a sum of an absolutely convergent series.
The hypothesis, the same that in \cite{11}, 
is slightly weaker than the hypothesis used in 
the mentioned previous works. 
We use, as in \cite{10} and \cite{11}, the formalism of 
an abstract polymer system.
Some recent developments of this approach can be found 
in ref.\ \cite{13}.

In the present note we follow refs.\ 
\cite{6} and \cite{12}.
The proofs, however, have been improved 
and simplified.

\medskip


\begin{thebibliography}{99}

\small
\parskip=0pt

\bibitem{1}
J.L. Lebowitz and O. Penrose,
Convergence of virial expansions,
J. Math. Phys.
5, 841--847 (1964) 

\bibitem{0}
J. Groeneveld,
Two theorems on classical many-particle systems, 
Phys. Letters 3, 50--51 (1962)

\bibitem{2}
J.L. Lebowitz,
Statistical mechanics --
a review of selected rigorous results,
Ann. Rev. Phys. Chem.
19, 389--418 (1968) 

\bibitem{3}
D. Ruelle,
Statistical mechanics: Rigorous results,
Benjamin, New York, 1969.

\bibitem{4}
B. Simon,
The statistical mechanics of lattice gases,
Princeton University Press, Princeton, 1993.

\bibitem{5} 
R.A. Minlos, Ya.G. Sinai,
The phenomenon of ``phase separation'' at low temperatures
in certain lattice models of a gas, I and II,
Math. USSR Sbornik 2, 339--395 (1967) and
Trans. Moscow Math. Soc. 19, 121--196 (1968)

\bibitem{6}
G. Gallavotti, A. Martin-Lof, S. Miracle-Sole,
Some problems connected with the description of
coexisting phases at low temperatures in Ising models,
in ``Mathematical Methods in Statistical Mechanics'', 
A. Lenard, ed., pp.\ 162--202, Springer, Berlin, 1973.

\bibitem{7}
E. Seiler,
Gauge theories as a problem of constructive quantum
field theory and statistical mechanics,
Springer, Berlin, 1982.

\bibitem{8}
D.C. Brydges,
A short course on cluster expansions,
in ``Critical Phenomena, Random Systems, Gauge Theories'',
K. Osterwalder and R. Stora, eds., pp.\ 129--183, 
Elsevier Science, Amsterdam, 1986.

\bibitem{9}
G. Gallavotti, 
Teoria dei polimeri, 
unpublished. 

\bibitem{10} 
R. Koteck\'y, D. Preiss,
Cluster expansion for abstract polymer systems,
Commun. Math. Phys.\ 103, 491--498 (1986)

\bibitem{11}
R.L. Dobrushin, 
Estimates of semi-invariants for the Ising
model at low temperatures,
in ``Topics in Statistical and Theoretical Physics: 
F.A. Berezin memorial volume'', 
R.L. Dobrushin et al., eds.,
pp.\ 59--81, 
Amer. Math. Soc., 1996.

\bibitem{12} 
S. Miracle-Sole,
On the convergence of cluster expansions, 
Physica A 279, 244--249 (2000)

\bibitem{13}
R. Fernandez, A. Procacci,
Cluster expansions for abstract polymer models. 
New bounds from an old approach. 
Commun. Math. Phys.\ 274, 123--140 (2007)
 
\end{thebibliography}
\end{document}